\documentclass[aps,prb,nofootinbib,twocolumn,superscriptaddress,showpacs]{revtex4-1}
\usepackage{graphicx}
\usepackage{bm}
\usepackage{color}
\usepackage{hyperref}
\usepackage{amsmath}
\usepackage{enumerate}
\usepackage[normalem]{ulem}

\begin{document}
\title{Energy and particle currents in a driven integrable system}
\author {D. Crivelli}
\affiliation{Institute of Physics, University of Silesia, 40-007 Katowice, Poland}
\author{P. Prelov\v{s}ek}
\affiliation{Faculty of Mathematics and Physics, University of Ljubljana, SI-1000 Ljubljana, Slovenia }
\affiliation{J. Stefan Institute, SI-1000 Ljubljana, Slovenia }
\author {M. Mierzejewski}
\affiliation{Institute of Physics, University of Silesia, 40-007 Katowice, Poland}

\begin{abstract}
We study the ratio of the energy and particle currents ($j^E/j^N$) in an integrable one dimensional system of interacting fermions.
Both currents are driven by a finite (nonzero)  dc electric field.  In doped insulators, where the local conserved quantities saturate the so called Mazur bound on the charge stiffness, $j^E/j^N$ agrees with the linear--response theory, even though such agreement may be violated for each current alone.  However, in the metallic regime with a non-saturated Mazur bound, the ratio $j^E/j^N$ in a driven system is shown to be much larger than predicted by the linear--response theory.     
\end{abstract}
% 72.10.-d Theory of electronic transport; scattering mechanisms
% 71.27.+a Strongly correlated electron systems; heavy fermions
% 72.10.Bg General formulation of transport theory

\maketitle
\section{Introduction and Motivation.}

The physics beyond the linear response (LR) regime is interesting for basic research and potentially important for the future applications.  The underlying phenomena  have recently become accessible to novel experimental techniques like ultrafast pump--probe spectroscopy of solid state systems or measurements of the relaxation processes in ultracold atoms driven far from equilibrium.      Significant progress has also been achieved in the theoretical description of solids driven by a finite (nonzero) electric field \cite{Amaricci2012,Arrachea2002,Aron2012,einhellinger2012,Joura2008,meisner,lev2011,lev2011_1,janez1,Boulat2008}. Recently developed numerical approaches allow to study response  to the electric field of (almost) arbitrary strength. In particular, applicability of the LR theory  has been tested for a weak--to--moderate driving  \cite{Russomanno2013,Mierzejewski2011,Steinigeweg2012,Karrasch2013}, while  for extremely strong fields one has studied  the Bloch oscillations in systems of strongly interacting carriers.\cite{Buchleitner2003,Eckstein2011,Freericks2008,Mierzejewski2010,covaci}  At the same time, the combined transport of energy and charge, which determines the thermoelectric properties, has been studied mostly  in the LR regime with only a few attempts to the nonequilibrium regime \cite{leijnse,sanchez2013,Ajisaka2012,ourcouple}.  Promising results concerning enhanced the thermoelectric performance have been reported for low--dimensional systems  \cite{Benenti2013,Dresselhaus2007,Dubi2011,Kim2009} for systems with ballistic (coherent) charge carriers \cite{Hlubek2010,Benenti2013a,Karlstroem2011} as well as for systems with strongly interacting electrons  \cite{Arsenault2013,Zlatic2012,Zlatic2014,Peterson2007,Shastry2009,Zemljic2005,Paul2003,Kargarian2013}.  

We first note that not all currents which are well established in the LR theory remain uniquely defined also in a generic nonequilibrium situation. 
Related to conservation laws and continuity equations, the energy and particle currents are well defined also beyond LR, while e.g. the heat current is not. 
The main objective of our research is to establish the ratio of the energy current $ j^E $ and the particle current  $ j^N $ 
\begin{align} 
R(t) = \frac{j^E(t)}{j^N(t)},
\label{eq:time-ratio}
\end{align}
in a homogeneous {\em integrable} system 
which at time $t=0$ is in equilibrium while for  $t>0$ is driven by a finite electric field $F$. 
Generic (nonintegrable) systems show a dissipative transport, hence a steady driving induces steady currents  
$j^{N(E)}(t \rightarrow \infty)=\mathrm{const}$ and the dc ratio $R(t \rightarrow \infty)$ is well defined.
One would wish to discuss directly the heat current usually expressed as $j^Q=j^E-\mu j^N$, but the chemical
potential $\mu$ is essentially an equilibrium concept.
While $j^E$ is still not the heat current, at least under close--to--equilibrium conditions $R(t \rightarrow \infty)$ can be related to various thermoelectric properties,\cite{Goupil2011} e.g. the Peltier coefficient $\Pi=R(t\rightarrow \infty)-\mu$.

Integrable systems display unusual relaxation \cite{gge,Eckstein2012,Cassidy2011,Steinigeweg2012}  and transport properties \cite{ZotosIdeal,Mierzejewski2010,Mierzejewski2011,Tomaz2011, Marko2011,Sirker2009,Steinigeweg2011}.
In particular they show a {\em ballistic} transport quantified by a nonzero charge stiffness leading to singular response functions.\cite{ZotosIdeal,Naef97,Heidrich-Meisner2003,Mukerjee2008,Sirker2009,Orignac2003,Steinigeweg2014} 
On the one hand, the basic understanding of the  ballistic transport is that a steady driving induces a steadily growing currents 
$j^{N(E)}(t ) \propto t$. On the other hand, in the tight--binding models the expectation values of currents cannot become arbitrarily large.
This poses limits on the time--window in which currents may indeed vary linearly in time. It has recently been shown for driven integrable systems that the particle
current undergoes the Bloch oscillations \cite{Mierzejewski2010} and it is straightforward to expect the same also for $j^E$.      
Note also that finite electric field acts as a source of the currents, but doubles as an integrability breaking mechanism.
Despite singular response functions, the dc ratio of energy and particle currents has been expected to remain well defined and finite \cite{Zemljic2005} at least in the 
LR regime ($F\rightarrow 0$). However, the above arguments indicate that it is by far not obvious whether/when/why it may actually take place under a finite driving.
It is the main problem which we address in this paper.

The manuscript is organized as follows. In the subsequent section we introduce a model and the specify the details of driving.   
Then, as a test of our approach we study a generic system and show how the LR results for $R(t)$ can be extracted 
from time evolved observables. Next we turn our approach on integrable systems, where strictly equilibrium predictions for the ratio of currents are ambiguous due to singularities 
of the response functions. First, we investigate a doped insulator, for which the ballistic transport can be explained as originating from local conserved quantities.\cite{Herbrych2011} 
Finally, we present conjectural results for a  metallic system in which a relation between charge stiffness and local conservation laws has not been established.

\section{Setup and Methods}
The system under study is a closed, homogeneous 
one--dimensional ring of charged,  spinless, but interacting fermions. The Hamiltonian is that of the \textit{t-V-W} model, arranged on a periodic ring of $L$ sites:
\begin{align}
H(t)=&-t_0 \sum_i \left[ {\mathrm e}^{i \phi(t)}\; c^{\dagger}_{i+1} \, c_i +{\mathrm H.c.} \right]  \nonumber \\
    & + V \sum_i \tilde{n}_i \, \tilde{n}_{i+1} + W \sum_i \tilde{n}_i \, \tilde n_{i+2},
\label{hamiltonian}
\end{align}
where $n_i =  c^{\dagger}_{i} \, c_i$, $\tilde n_i=n_i -1/2$
and  $t_0$ is the hopping integral. $V$ and $W$ are repulsive interactions between first and second nearest neighbors, respectively. The latter interaction is introduced to break integrability in a controlled manner and to allow for the normal diffusion (at least at weak driving).  
Below we use units in which $\hbar=k_B=t_0=1$. 

%and incorporated in the time-dependent Hamiltonian  through Peierls substitution. 

The dynamics is studied by explicitly solving the time-dependent Schr\"odinger equation for a pure quantum state. 
The initial equilibrium state $|\Psi(t=0) \rangle$ is determined
from the Microcanonical Lanczos Method (MCLM) \cite{Long2003}
for the energy $E_0= \langle \Psi(0)| H(0) |\Psi(0) \rangle$ corresponding to a target inverse temperature $\beta$.
If not specified otherwise we take $\beta \simeq 0.4$, while a typical 
energy uncertainty is $\delta E_0=\langle \Psi(0)| [H(0)-E_0]^2|\Psi(0) \rangle^{1/2} \simeq 0.003$. 
For time $t>0$ the system is driven by a constant electrical field $F$ \cite{ring1,ring2,ring3,ring4,Arrachea2002}, induced by linearly varying magnetic flux $\phi(t) = - F t$. The evolution under driving $|\Psi(0)\rangle \to |\Psi(t)\rangle$ is obtained by means of a fourth order expansion \cite{Alvermann2012} of the time ordered exponential, with Chebyshev approximation of the unitary propagators \cite{chebytime} on successive small time intervals. The evolution is thus unitary and numerically accurate, allowing long timescales up to $t \alt 1 / \delta E_0$.

\begin{figure}[tb]
\includegraphics[width=0.45\textwidth]{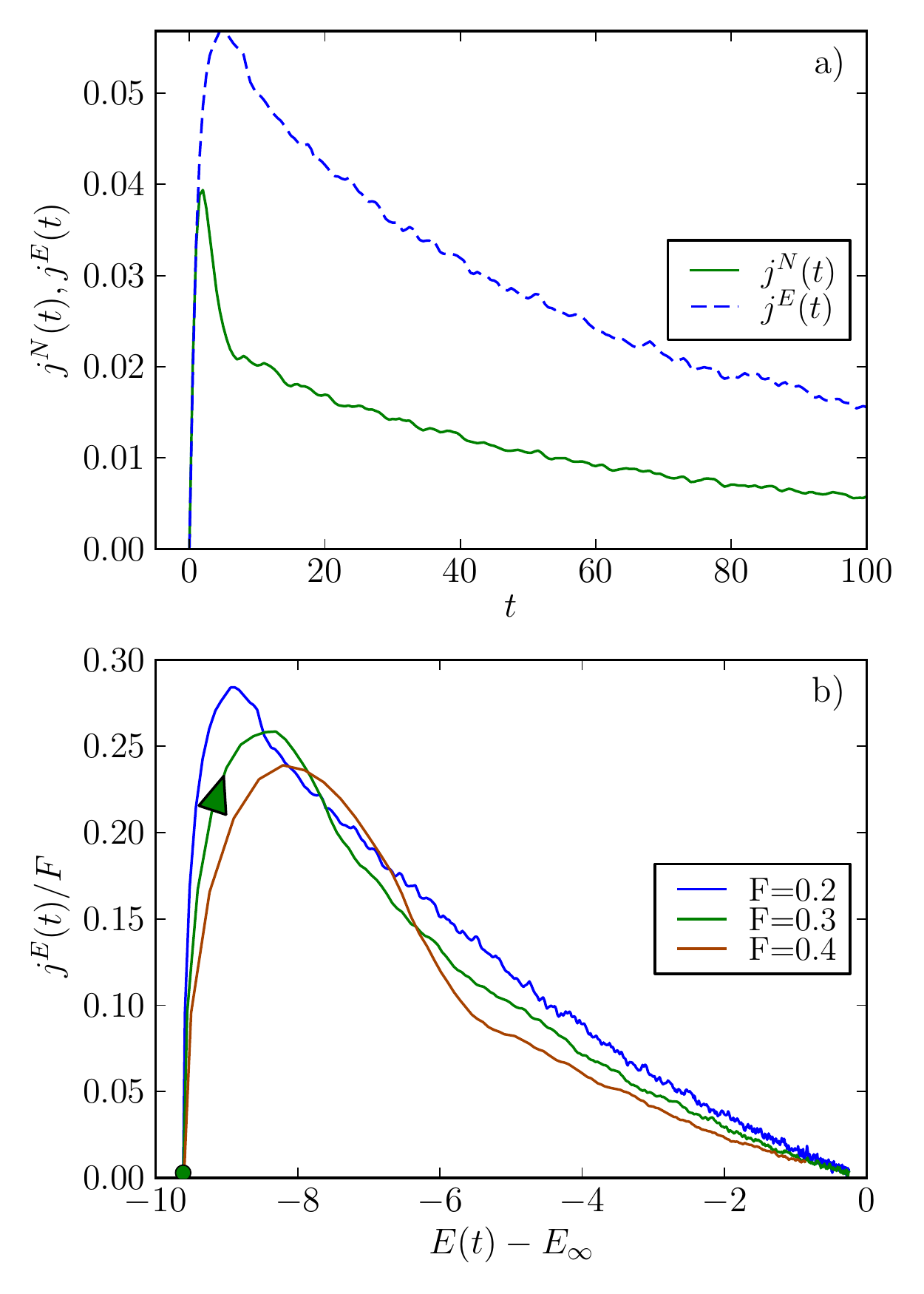}
\caption{(Color online) Results for a generic system with $V=3, W=1, L=24, N=10$. (a) Time dependence of energy and particle 
currents for $F=0.2$.
%, which for long times are proportional to each other 
(b) Ratio $j^E/F$ for different fields $F$ as function of the instantaneous energy $E$ relative to the energy at infinite
temperature $E_{\infty}$. Point marks the initial state and arrow shows the direction of time--evolution.
}\label{fig1}
\end{figure}

We study the particle (charge) current $j^N = \langle J^N \rangle$ and the energy current $j^E = \langle J^E \rangle$, both induced by the same field $F$. The currents 
follow uniquely from the continuity relations for the local charge and energy densities \cite{Mierzejewski2013,Naef97, Mukerjee2008} 
and have the form:
\begin{align}
J^N(t)=& \frac{1}{L} \sum_i J^N_i =  \frac{1}{L} \sum_i \left[i \mathrm{e}^{i\phi} c^{\dagger}_{i+1}\, c_{i}+ \mathrm{H.c.}\right],
\label{jn} \\
J^E(t) =&  \frac{1}{L} \sum_i J^E_i = \,-\frac{1}{L}\sum_i \Big\lbrace[i \mathrm{e}^{2i\phi} c^{\dagger}_{i+1}\, c_{i-1}+{\mathrm H.c.}] \nonumber \\
  + \frac{J^N_i}{2} \big[ 3W & (\tilde{n}_{i+3}\!+\!\tilde{n}_{i-2})+ (2V\!-\!W)(\tilde{n}_{i+2} \!
 + \! \tilde{n}_{i-1}) \big] \Big\rbrace.
 \label{je}
\end{align}
The equilibrium continuity equation for charge holds true also in driven systems, because driving does not influence the conservation of particles.
However, the energy of a driven system is not conserved. Therefore, the relevant continuity equation contains also the source terms 
which for systems driven by electric field represent the effects of the Joule heating:
\begin{equation}
\frac{\mathrm d}{{\mathrm d} t}  
\langle H_i \rangle +
\nabla \langle  J^E_i   \rangle   = F \langle   J^N_i \rangle.  
\end{equation}
Here, $H_i$ is the energy density operator, $H=\sum_i H_i$. In the LR regime the currents  can be equivalently derived from the polarization operators  \cite{Shastry2009, Louis2003, Paul2003}.

Departure from half--filling ($N \ne L/2$) is necessary to obtain nonzero  $j^E$ since at half filling 
the Hamiltonian is invariant under the particle--hole transformation $c_i \rightarrow (-1)^i c^{\dagger}_i$  while
$J^E \rightarrow -J^E$  under this transformation. Furtheron, the number of charged fermions is taken to be $N=10$ for the $L=24$ site ring or $N=9$ for $L=26$ sites 
both slightly below half--filling. We investigate systems with $V=1.5$ and $V=3$ which for $W=0$ correspond, respectively, to doped metals and insulators \cite{Rigol-tVW}.
The latter insulating phase is induced by a short range fermion--fermion interaction ($V$) and is charge ordered.  
Hence, it shares common properties with Mott insulators as well as with charge density wave insulators.

\section{Generic response of nonintegrable systems} 
\label{Sec:Nonintegrable}
Figure \ref{fig1}(a) shows the time--dependence of both currents in a driven generic system. 
Shortly after turning on the electric field, $j^{E(N)}$ can be easily determined from the equations of motion \cite{Mierzejewski2010}
\begin{equation}
\frac{d}{dt}  j^{N(E)}   =  -\tau^{N(E)} \dot{\phi} + i \langle [H,J^{N(E)}] \rangle,
\label{eq_motion}
\end{equation}
where $ \tau^{N} = -\langle \partial_{\phi} J^{N} \rangle $ and $\tau^{E} = -\langle \partial_{\phi} J^{E} \rangle $ are stress coefficients (tensors in general)
determining the short--time LR to the flux change. 
The last term in Eq. (\ref{eq_motion}) vanishes for the initial equilibrium state, hence the short--time ratio of the energy and particle currents
\begin{align}
R(t\rightarrow 0^{+}) = \frac{\tau^{E}}{\tau^{N}}, 
\label{eq:initial-ratio}
\end{align} 
is field--independent and always consistent with the LR theory \cite{Louis2003,Zemljic2005}. 

%where $\tau^{N} = H_{\mathrm{kin}}/{L}$ (Eq. \ref{kinetic} defines $H_{\mathrm{kin}}$) in particular. 
%\begin{align}
%\tau^{N} =& =  -\frac{1}{L} \sum_i \left[c^{\dagger}_{i+1}\, c_{i}+ \mathrm{h.c.}\right] = H_{\mathrm{kin}} / L,
%\label{eq:tau_nn} \\
%\tau^{E} =& \,\frac{1}{L}\sum_i \Big\lbrace[c^{\dagger}_{i+1}\, c_{i-1}+{\mathrm h.c.}] \\
%  + \frac{\tau^{N}_i}{2} \big[ 3W & (\tilde{n}_{i+3}+\tilde{n}_{i-2})+ (2V - W)(\tilde{n}_{i+2} 
% +\tilde{n}_{i-1}) \big] \Big\rbrace .
% \label{eq:tau_ne}
%\end{align}

\begin{figure}[tb]
\includegraphics[width=0.45\textwidth]{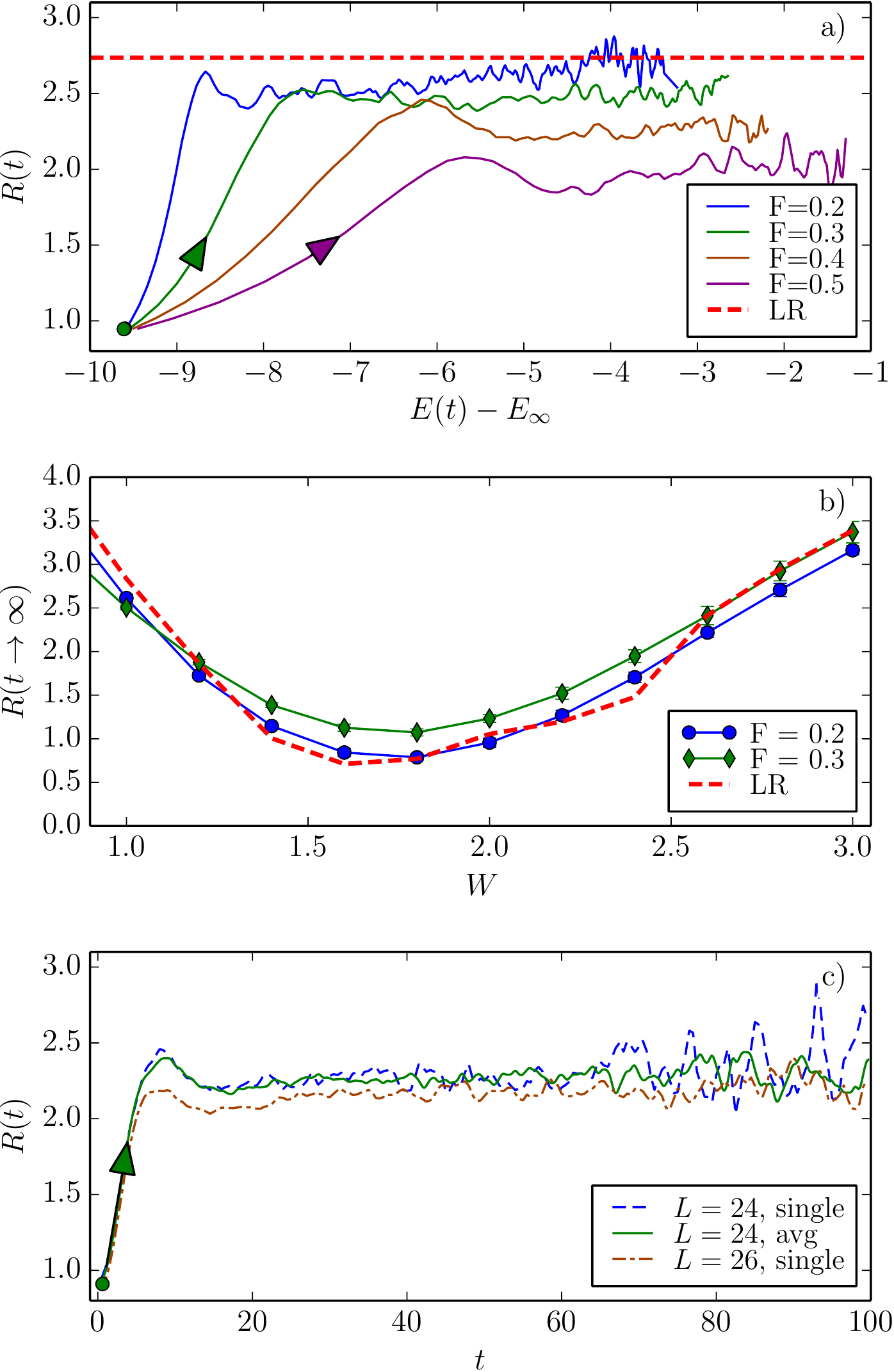}
\caption{(Color online) Results for a generic system with $V=3$, $L=24$, $N=10$.
(a) Ratio $R=j^E/ J^N$ for $W=1$ as a function of the instantaneous energy compared to the LR results.
%converging for vanishing $F$ to the LR result from Kubo formula 
(b) Long--time ratio $R(t\rightarrow \infty)$.
The LR results have been computed at the average temperature of evolution ($\beta\simeq 0.05$).
(c) Time dependence of $R(t)$ for $V=3, W=1, F=0.4$ with a single initial state (dashed line), averaged over 16 initial states (solid line), and for a bigger system ($L=26, N=11$) with similar concentration of fermions (dot dashed line).
}\label{fig2}
\end{figure}

In a closed tight--binding model, constant $F$ cannot induce strictly time--independent current since such d.c. response
would cause a steady and unlimited in time increase of the energy\cite{Russomanno2013}, while the energy spectrum is bounded from above.
However, the long--time dependence can still be reconciled with LR theory\cite{Mierzejewski2010,Mierzejewski2011,Eckstein2011}
provided these nonlinear effects of heating are properly filtered out. 
For a weak but finite $F$ the system undergoes a quasiequilibrium evolution, when the instantaneous expectation 
values of observable are uniquely determined only by  $F$ and the instantaneous energy E(t)
or (equivalently) by the instantaneous effective 
temperature.\cite{Mierzejewski2013} 
Consequently one should also consider the dc response functions
%as function of the effective temperature of the system, 
%by the relation $T_{\text{eff}} = T[E(t)]$ which 
as quantities which depend 
on $E(t)$.
An extended form of LR, $j^{N(E)}(t) \simeq \sigma^{N(E)}[\omega\to 0, E(t)] F$ holds true 
in the quasi--equilibrium regime.\cite{Mierzejewski2010} In this regime the ratios $j^{N(E)}/F$ weakly depend on $F$ 
and vanish \cite{Eckstein2011,Mierzejewski2010} when the system's energy approaches its value 
at the infinite temperature, $E_{\infty}$.
Both results are  explicitly shown in Fig. \ref{fig1}(b) for the case of $j^{E}$.
For $T \rightarrow \infty $ the energy dependence of the response functions cancels out and $R(t\rightarrow \infty)$ should be a well defined 
and finite. Figure \ref{fig2}(a) shows that it is actually the case. Moreover, the results obtained from the time--evolution remain in a good agreement with the LR results 
for the high--temperature regime:
\begin{align}
R(t\rightarrow \infty) \simeq
\left.
\frac{\sigma^{E}(\omega \to 0, E) }
     {\sigma^{N}(\omega \to 0, E) } \right|_{E\rightarrow E_{\infty}} .
\label{eq:lr-ratio}
\end{align}
For a nonintegrable systems at nonzero temperature $\sigma^{N}(\omega,E)$ and  $\sigma^{E}(\omega,E)$  
are regular. In the MCLM they are proportional to the current-current correlator on the state $|\Psi_{\beta}\rangle$ for the energy $E_{\beta}$ corresponding to the inverse temperature $\beta$:
\begin{align}
\sigma^{N(E)}_{\text{reg}}(\omega) =& 
L \frac{1 - e^{-\beta\omega}}{\omega} \; \mathrm{Im}\, C^{>}_{N(E)}(\omega), \label{eq:conductivity}\\
 C^{>}_{N(E)}(\omega) =& \langle \Psi_\beta | J^{N(E)}  (\omega^{+} +E_{\beta} - H)^{-1}  J^N | \Psi_\beta\rangle. \label{eq:correlator}
%
% \\ =& \pi \sum_n^{N_L}
%\overline{\langle  \tilde{\psi}_n | J^X \psi_0 \rangle}
%\langle \tilde{\psi}_n | J^N \psi_0\rangle
%\delta_{\eta}(\omega +\epsilon_0 - \tilde{\epsilon}_n).
\end{align}
We use a Lanczos expansion with Lorentzian broadening $\omega^{+} = \omega + i 0 ^{+}$. 
%, with $\tilde{\psi}_n$ the $n$-th approximate eigenvector in the Krylov basis, $\tilde{\epsilon}_n$ the corresponding approximate eigenvalue, delta functions are broadened with Lorentzians of width $\eta$, and $|\psi_0\rangle$ are the microcanonical states with energy $\epsilon_0$, where the temperature dependence is implicit in the choice of $\epsilon_0$. Analogous expressions hold substituting $j^N$ for $j^E$.

In a driven system, the estimate of $R$ can be obtained robustly from the least-squares scaling of 
$j^N$ against $j^E$ for long times ($t>t_0 \simeq 50$)
\begin{align}
\frac{d}{d R}\sum_{t_i > t_0} \left[j^E(t_i) - R\, j^N(t_i)\right]^2 = 0.
\label{eq:least-squares-ratio}
\end{align} 
The results are shown in Fig. \ref{fig2}(b) for $V=3$ and various $W$. The data obtained for  driven systems nicely recover the equilibrium results from the standard LR approach.
Extracting the LR limit is thus possible from the time-dependent quantities, since the ratio of Eq. (\ref{eq:time-ratio}) is a well behaved monotonic function of $F$. This holds true as long as the driving is not as strong as to induce the Bloch Oscillations (BO) of the currents \cite{Freericks2008,Eckstein2011,Mierzejewski2011}, which eventually occur also in generic nonintegrable systems.

The tiny oscillations of currents around their average values (see Fig. \ref{fig1}(b)) originate from the fact that we carry out calculations for a finite quantum system and for a single initial state. However  small are these oscillations they eventually dominate when the system approaches $\beta\to 0$ and the smooth components of the currents vanish. 
Then, the numerical results for $R(t)$ being the ratio of two vanishing quantities unavoidably becomes  noisy (see Fig. \ref{fig2}(a)).
These oscillations have no physical meaning and can be reduced by either increasing the system size or by averaging over many initial states. Both cases are shown in Fig. \ref{fig2}(c).

\section{Doped integrable insulator}
After showing that our method reliably applies to the generic case, furtheron we restrict the scope to driven integrable systems and set $W = 0$.
In equilibrium the real part of the dynamical conductivity has two separate contributions:
\begin{align}
\sigma^{N}(\omega) = 2\pi D^{N} \delta(\omega) + \sigma^{N}_{\text{reg}}(\omega).
\label{eq:total-conductivity}
\end{align}
The regular part $\sigma^{N}_{\text{reg}}$ is connected with normal (diffusive) behavior while the singular one is
 weighted by the stiffness $D^{N}$ and implies anomalous (ballistic) transport as well as non-decaying currents. 
The sum rule $\int_{-\infty}^{\infty} \sigma^{N}(\omega) d\,\omega = \pi \tau^{N}$ allows to normalize and weight the different contributions
with the previously defined $\tau^{N}$ operator, thus linking initial-time [see Eq. (\ref{eq_motion})] with the  dynamical response.
Since $j^E$ is conserved, the regular part  of $\sigma^{E}(\omega)$ vanishes and the LR response of the energy current 
is purely singular
\begin{equation}
\sigma^{E}(\omega)=2\pi D^{E} \delta(\omega)= \pi \tau^{E} \delta(\omega).
\label{sne}
\end{equation}

In the case of doped insulators ($V>2$) 
the Drude   weight $D^{N}$ can be well estimated from the Mazur bound by taking the overlap of $J^N$  
with  a single conserved quantity - the energy current 
\cite{Herbrych2011,Mukerjee2008,Naef97,ZotosIdeal}:
\begin{align}
D^{N} \approx D^{N}_{\text{Mazur}} =   \frac{\beta L}{2}\; \frac{\langle J^N\, J^E \rangle^2}{\langle J^E\, J^E \rangle^{\phantom{2}}}
\label{eq-mazur}.
\end{align}

\begin{figure}
\includegraphics[width=0.45\textwidth]{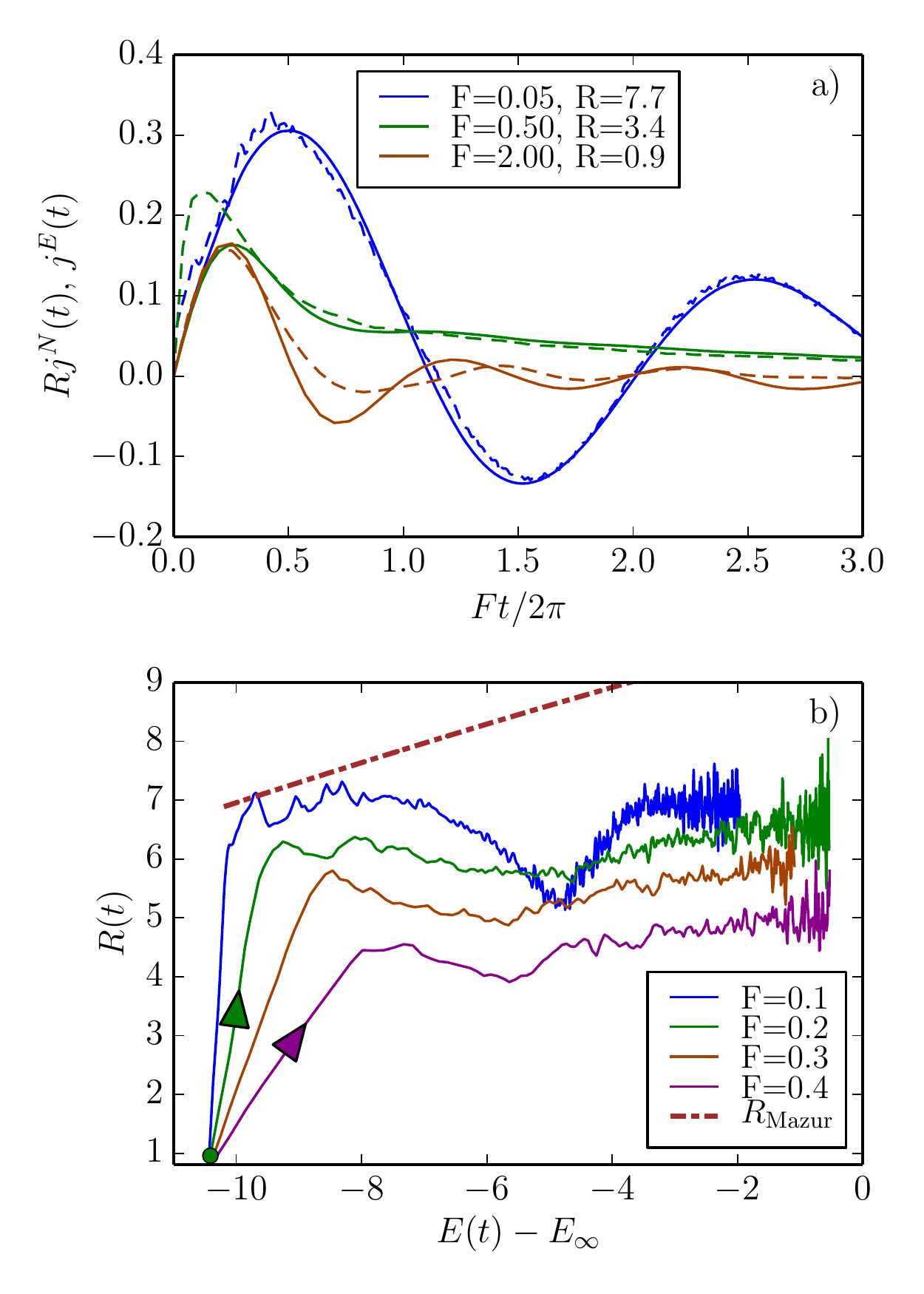}
\caption{(Color online)  Currents for $L=24,N=10$, $V=3$ and $W=0$. (a) $j^E(t)$ (solid lines) together with $R j^N(t)$ (dashed lines) for $R$ shown in the legend. (b) $R(t)$ vs instantaneous energy compared to $R_{\text{Mazur}}$ [see Eq. (\ref{eq-mazur-bound})].}
\label{fig3}
\end{figure}

\begin{figure}
\includegraphics[width=0.45\textwidth]{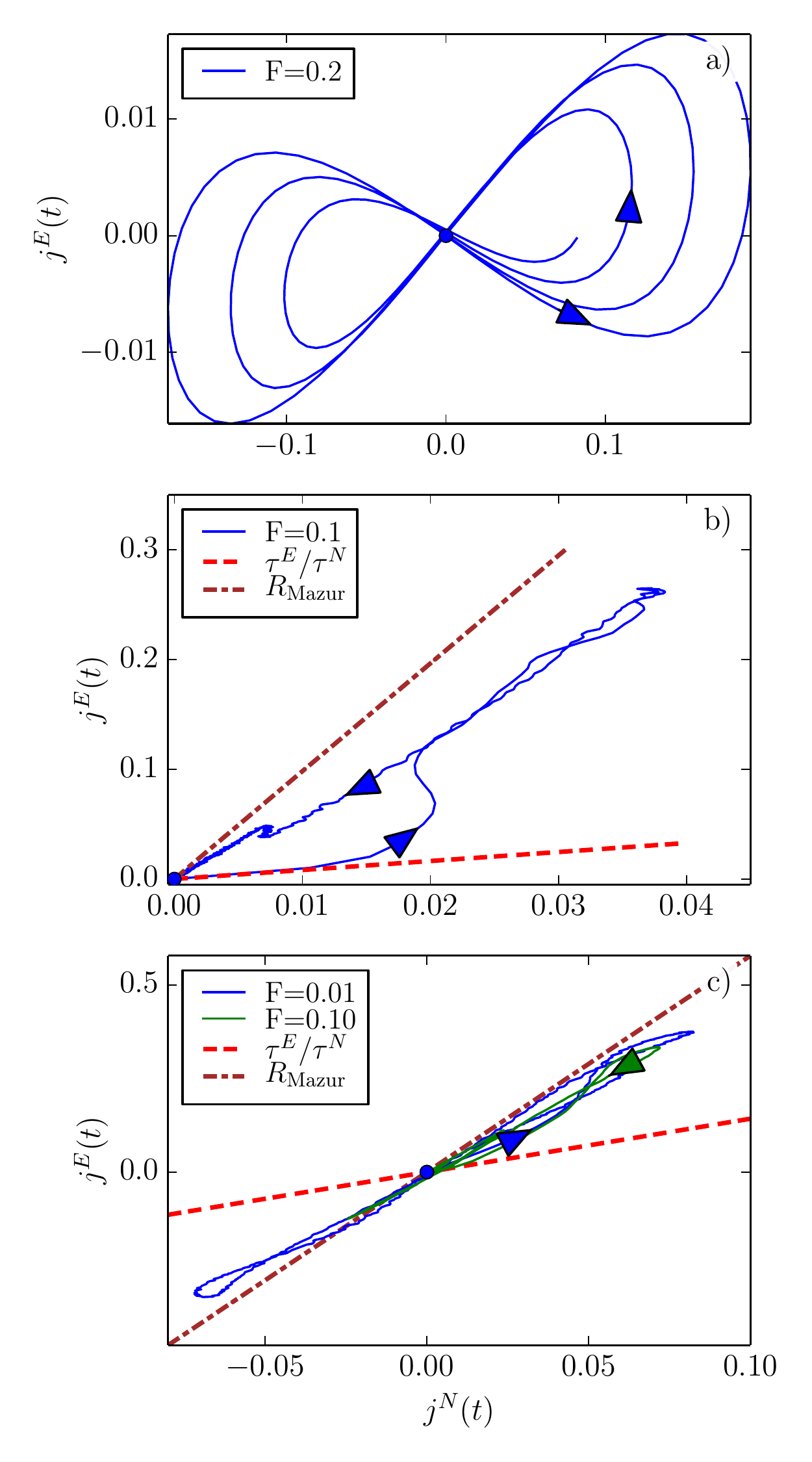}
\caption{ (Color online) Parametric plots $j^E(t)$ vs. $j^N(t)$ for $W=0$. 
 (a) Results for  $L=24,N=10$  and extremely weak interaction $V=0.2$ when  
 $j^E(t)$ oscillates with frequency twice larger than $j^{N}(t)$.  
(b) The same as in (a) but for $V=3$ while 
$L=26,N=9$ and $V=3$ are used in (c).
In the two latter panels the straight lines 
show $\tau^E/\tau^N$ for the initial $\beta$ [see Eq. (\ref{eq:initial-ratio})] and  $R_{\text{Mazur}}$ for $\beta \rightarrow 0$ [see Eq. (\ref{eq-mazur-bound})].}
\label{fig4}
\end{figure}

According to the LR theory, $j^N$ and $j^E$ should grow linearly in time for a dc driving. 
However, this linear growth cannot be unlimited in time under a finite driving as argued in the preceding sections. 
Then, the currents may develop either into BO \cite{Freericks2008,Eckstein2011} or into quasistatic current as observed for generic systems. The latter is also possible 
since finite $F$ breaks the integrability. Figure \ref{fig3}(a) shows that the strength of driving determines the
scenario which prevails.  We observe oscillatory response in the limits of very weak and very strong driving, 
and quasisteady currents for the intermediate $F$.
  
The relation between $j^N$ and $j^E$ can be inferred from Fig. \ref{fig3} as well as from the parametric plots
shown in  Fig. \ref{fig4}. For a weak--to--moderate driving both currents are roughly proportional to each other. It holds true   independently of whether these currents are quasistatic as shown in
Figs. \ref{fig3}(b) and \ref{fig4}(b) or undergo the BO (Fig. \ref{fig4}(c)). Hence in this regime the ratio $R(t)$ is indeed well defined and meaningful despite the singular LR of the integrable system.  
The proportionality between oscillating currents $j^N$ and $j^E$ for $F \rightarrow 0 $ is rather unexpected. Such proportionality is evidently broken for BO under large $F$ 
(see Fig. \ref{fig3}(a) for $F=2$)  and/or for very weak $V$. Due to an exact doubling of the frequency of their oscillations  in the latter case  [see Eqs. (\ref{jn}),(\ref{je})] the currents form a damped Lissajous figures in the parametric plane ($j^E$,$j^N$) as shown in Fig. \ref{fig4}(a).

%\begin{figure}
%\includegraphics[width=0.45\textwidth]{fig5}
%\caption{(Color online)  New panels in previous plots
%}\label{fig5}
%\end{figure}

\begin{figure}
\includegraphics[width=0.45\textwidth]{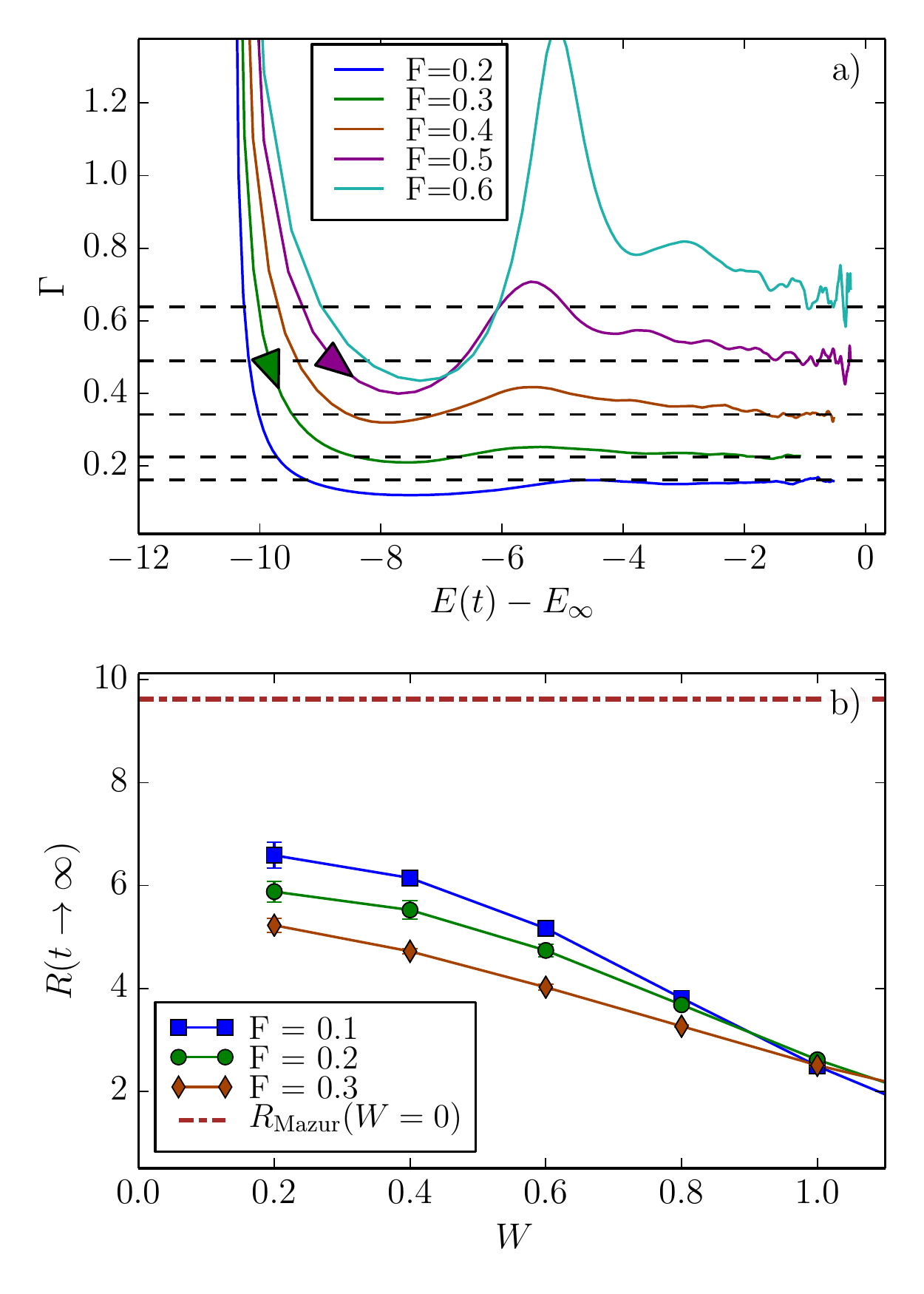}
\caption{(Color online) Results for $V=3$, $L=24$ and $N=10$. (a) Phenomenological scattering rate $\Gamma$ 
as a function of the instantaneous energy for $W=0$.
(b) $R(t\rightarrow \infty)$ for small but finite integrability-breaking interaction $W$.
The value of $R_{\text{Mazur}}$ (at $\beta \rightarrow 0$) for $W=0$ is shown for comparison.}
\label{fig5}
\end{figure}

In order to explain the numerical results we first focus on the regime of intermediate driving, when currents show the same steady behavior 
as in generic systems under quasiequilibrium evolution. Hence,  we apply a similar phenomenological modification of LR which turned 
out to be successful in the case of generic systems.\cite{Mierzejewski2010,Mierzejewski2011}  
Since the driving itself is sufficient to damp oscillations of the energy current, the main effects must be the  broadening of the singular response functions  \cite{Alvarez2002}.
A phenomenological attempt would be to modify Eq. (\ref{sne}) using a Lorentzian ansatz with an effective scattering rate $\Gamma$
\begin{align}
\delta(\omega) \longrightarrow \frac{1}{\pi} \frac{\Gamma}{(\omega^2 + \Gamma^2)}.
\end{align}
It leads to an effective dc response $\sigma^{E}(\omega\to0) = \tau^{E} / \Gamma$ and a quasistatic energy current
\begin{align}
j^E = \frac{\tau^{E}}{\Gamma} F.
\label{modlr}
\end{align}
We have used this formula together with the numerical data for $j^E(t)$ and determined the (phenomenological) effective scattering rate shown in Fig. \ref{fig5}(a). 
One may observe that  $\Gamma$ increases with $F$ and  
after the initial transient it becomes independent of the instantaneous energy.
Therefore the heating effect (dependence on the energy)  is included entirely in the sum rule $\tau^{E}$, while $\Gamma$ 
describes solely the broadening of the response--function by external driving.  

It is also interesting that the numerical values of $\Gamma$ are very close to $F$. Hence the effective scattering (damping) rate 
is close to the frequency of the BO ($\omega_B=F$). 
Therefore, within this phenomenological picture the regime of the quasistatic current is just at the boundary  of overdamped BO.

The same reasoning should also hold for the particle current, however the numerical analysis would be much more demanding since close to half-filling  ($ \langle n \rangle \sim 1/2$) the stiffness
$D^{N} \ll \tau^{N}/2$  in contrast to $D^{E} = \tau^{E}/2$.  However, assuming that a single scattering rate gives the broadening of both
response functions, one may estimate the ratio $R(t\rightarrow \infty)$ in the quasiequilibrium regime 
\begin{align}
R(t \rightarrow \infty) =   \frac{D^{E}}{D^{N}} \simeq 
R_{\text{Mazur}}=  \frac{\tau^{E}}{\beta L}\frac{\langle J^E\, J^E \rangle^{\phantom{2}}}{\langle J^N\, J^E \rangle^2}.
\label{eq-mazur-bound}
\end{align}
Results in Fig. \ref{fig3}(b)  and  \ref{fig5}(b) show that  
$R(t \rightarrow \infty)$ is reasonably close  to $R_{\text{Mazur}}$, provided $F$ is small enough.  The averages at the rhs of Eq. (\ref{eq-mazur-bound}) were computed by means of the kernel polynomial method\cite{Weiße2006} 
in the canonical ensemble at the temperature determined by the instantaneous energy during the evolution.
% It is a regularized expansion in polynomials of the thermal weight, which are sampled stochastically over the Hilbert space, yielding accurate expectation values over a wide range of temperatures. 
The deviations between the results from the real--time dynamics and Eq. (\ref{eq-mazur-bound}) in  Fig. \ref{fig5}(b) are overestimated since 
the real--time currents are determined at $E(t) < E_{\infty}$ while $R_{\text{Mazur}}$ for  $E \rightarrow E_{\infty}$.

Quite surprisingly, the prediction (\ref{eq-mazur-bound}) is accurately fulfilled also for weaker driving when both currents oscillate.  
In Fig. \ref{fig4}(c) such behavior is shown for a different filling factor, providing an independent test.
After a short transient, the currents oscillate perfectly in phase with a relative amplitude $R$ satisfying the Mazur bound of Eq. (\ref{eq-mazur-bound}), regardless of $F$. 
This agreement makes a clear connection between the BO under finite but weak $F$ and the stiffnesses within the LR theory.  
Note also that this relation is broken for large $F$, when BO are independent of integrability and occur also in generic systems.

\begin{figure}
\includegraphics[width=0.45\textwidth]{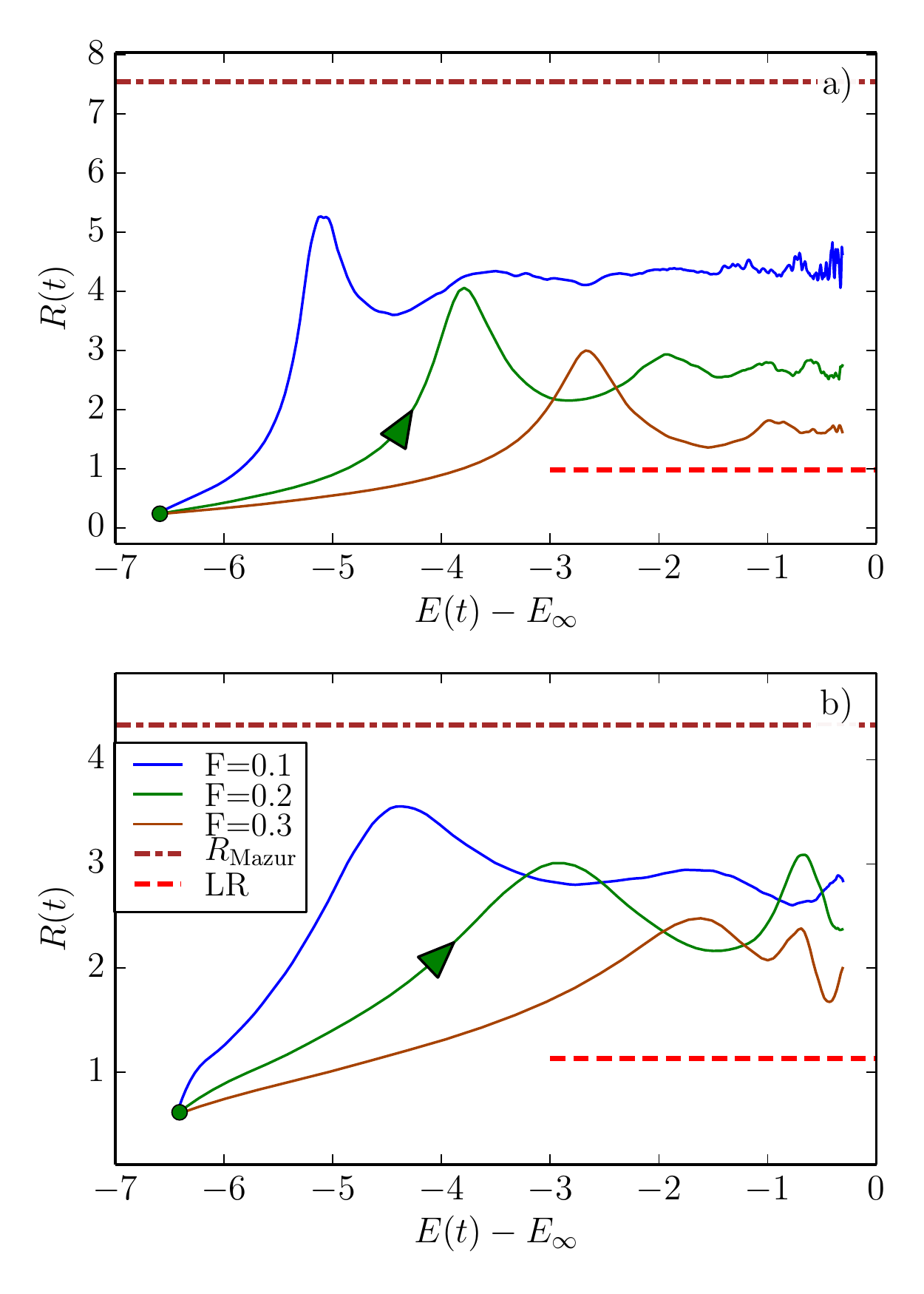}
\caption{(Color online) $R(t)$ as a function of instantaneous energy for $V=1.5$, $W=0$. (a) Results for  $L=24$ and $N=10$. (b)  $L=26$ and $N=9$. Dashed lines show $R_{\text{Mazur}}$  [see Eq. (\ref{eq-mazur-bound})] and the LR ratio $\frac{D^{E}}{D^{N}}$ [see Eq. (\ref{dtrue})] both at $\beta \rightarrow 0$.}
\label{fig6}
\end{figure}

\begin{figure}
\includegraphics[width=0.45\textwidth]{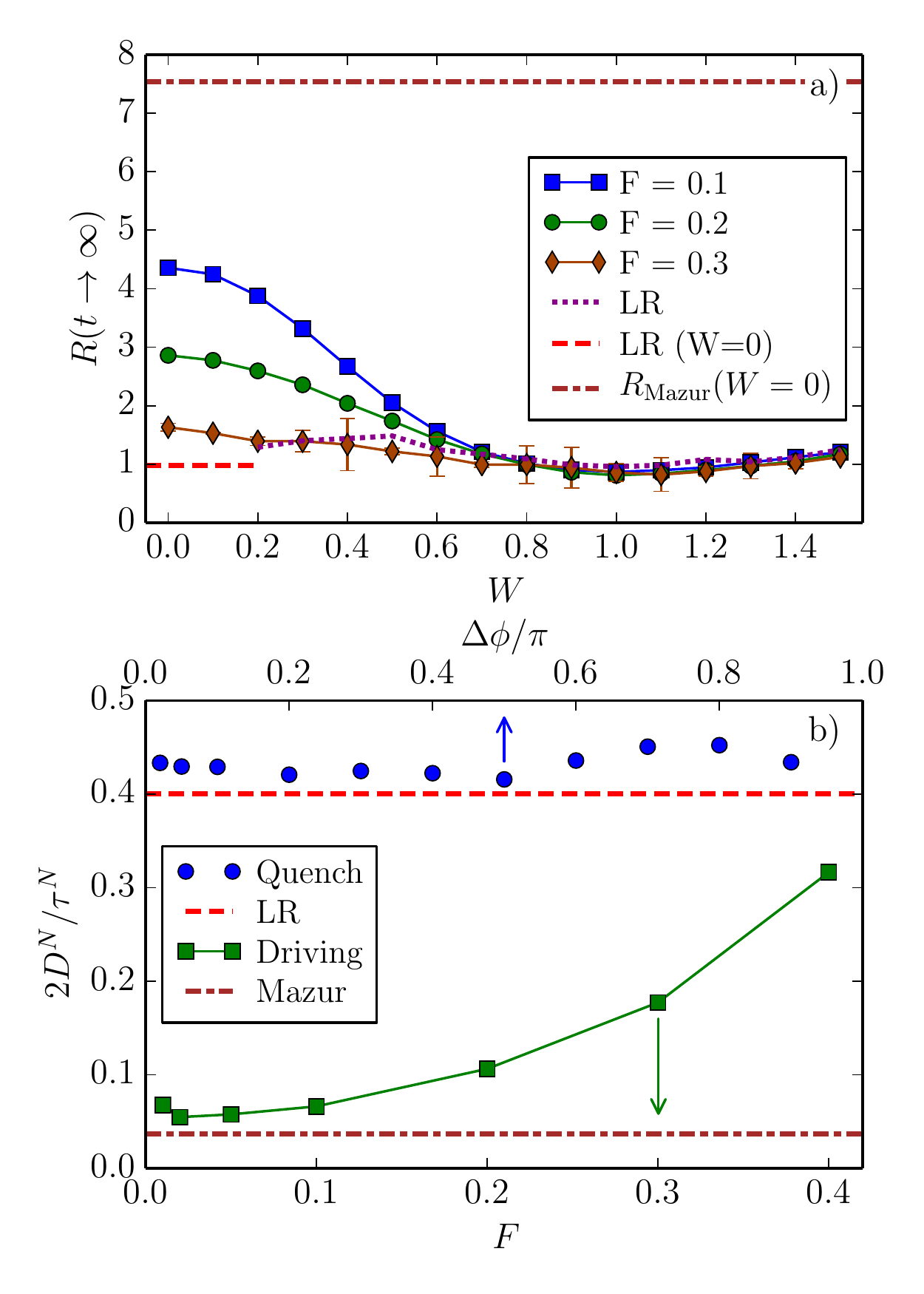}
\caption{(Color online) Results for $V=1.5$, $L=24,N=10$.
(a) $R(t\rightarrow \infty)$ for decreasing $W$ compared with the LR result [Eq. (\ref{eq:lr-ratio})]. For the case $W=0$ we also show $R_{\text{Mazur}}$ [Eq. (\ref{eq-mazur-bound})]
and LR ratio $\frac{D^{E}}{D^{N}}$ [see Eq. (\ref{dtrue})] both at $\beta \rightarrow 0$.
(b) The stiffness $D^{N}$, $D^{N}_{\text{Mazur}}$, $D^{N}_{\text{driving}}$ and $D^{N}_{\text{quench}}$ normalized to $\tau^{N}/2$ for $W=0$ as detailed in the text. 
}\label{fig7}
\end{figure}

\section{Integrable metals close to half-filling} 
We now turn to the case $V<2$ when the system is metallic at
arbitrary filling factor.  For moderate fields, currents again display only modest oscillations, so the ratio $R(t)$ can be determined directly (see Fig. \ref{fig6}).

It has been shown for integrable metals  at half--filling ($\langle n \rangle =1/2$) that the Mazur bound formulated in terms of strictly local
conserved operators fails, in particular  $D^{N}_{\text{Mazur}}=0$ while $D^N$ stays nonzero.  In order to saturate the Mazur bound, 
one (probably) needs to introduce quasi--local conserved operators.\cite{quasi11,quasi13,Ilievski2013} 
For slightly smaller concentration of fermions\cite{Herbrych2011} ($\langle n \rangle  < 1/2$),  $D^{N}$ is still much larger than  $D^{N}_{\text{Mazur}} $,  
hence the ratio $R(t \rightarrow \infty)$ was expected to be consistently lower than 
$R_{\text{Mazur}}$ given by (\ref{eq-mazur-bound}). However, the numerical data in Fig. \ref{fig6} show that $R(t)$ departures from LR and approaches $R_{\text{Mazur}}$, as if the energy current were the relevant conserved quantity. 
Fig. \ref{fig7}(a) shows $R(t \rightarrow \infty)$ calculated for small but nonzero $W$ in comparison to the LR results obtained directly from
the response functions as well as with $R_{\text{Mazur}} (W=0)$ given by  Eq. (\ref{eq-mazur-bound}). 
Upon decreasing $W$ one again observes that results for driven system departure from the predictions of LR theory towards $R_{\text{Mazur}}$ for $W=0$.
%The response of a driven integrable system splits for different fields with a ratio strongly enhanced for vanishingly small fields. $R(t \rightarrow \infty)$ reaches maximum for weak $F$ and small $W$ when both currents oscillate in phase in analogy to the integrable insulators.

We expect that breaking the integrability by finite $F$ is responsible for the observed departure from LR regime. 
In order to verify this expectation we have compared the response of the system driven by $F>0$ with its nonequilibrium relaxation at $F=0$.
In particular, we have calculated
$D^{N}_{\text{Mazur}}$  given by Eq. (\ref{eq-mazur}) as well as the actual charge stiffness calculated from the sum rule in Eq. (\ref{eq:total-conductivity}) 
taking the regular conductivity [Eq. (\ref{eq:conductivity})] in the initial MCLM state 
\begin{align}
D^{N} = \frac{\tau^{N}}{2}-\frac{1}{2\pi}\int_{-\infty}^{\infty} \sigma_{\text{reg}}^{N}(\omega)\, d\omega.
\label{dtrue}
\end{align}
These equilibrium results have been compared with two nonequilibrium cases. 
For a system evolving under finite $F$ one can estimate the charge stiffness from $R(t\rightarrow \infty)$ assuming that $j^E/j^N \simeq D^{E}/D^{N}_{\text{driving}}$ holds in long--time regime similarly to the case of doped insulators. Then,
\begin{align}
D^{N}_{\text{driving}} =  \frac{\tau^{E}}{2R(t\to\infty)}.
\end{align}
Finally, we have studied an instantaneous change of the magnetic flux which should also be consisted with LR.  
At $t=0$ we quench the flux $\phi(t) = \Delta \phi\, \theta(t) $  inducing an electric field $F(t) = -\Delta \phi\, \delta(t)$.
To the first order in $\Delta \phi$  the time--dependent particle current reads
\begin{align}
j^N(t) \!=-\tau^{N} \, \Delta \phi  - i L \int_{0}^{t} \langle [J^N(t'),J^N(t)] \rangle \Delta \phi\, dt'
\end{align}
which gives the peak value $j^N(t \rightarrow 0^+)  = -\tau^{N} \Delta \phi$ since the integrand is smooth. 
The real-time LR current  is given by $j^N(t) = \int \,d\omega\,F(\omega) \boldsymbol{\sigma}^{N}(\omega) e^{-i \omega^{+} t}$ where $F(\omega) = -\frac{\Delta \phi}{2 \pi}$ and $\boldsymbol{\sigma}^{N}$ is the complex conductivity. 
The regular part of $\boldsymbol{\sigma}^{N}$  is smooth and gives no contribution to $j^N(t)$ for $t \rightarrow \infty$.
%, as it averages to zero due to the Riemann-Lebesgue lemma:$\lim_{t\to\infty} \int_{-\infty}^{\infty} \boldsymbol{\sigma}_{\text{reg}}^{N}(\omega) e^{-i \omega t} \, d\omega = 0$. 
With the complex singular part $\boldsymbol{\sigma}^{N}_{\text{sing}}(\omega) = \frac{2iD^{N}}{\omega + i 0^{+}}$, the current after the quench stabilizes to
\begin{align}
j^N(t \to \infty)\! =\! - \int_{-\infty}^{\infty} \frac{\Delta \phi}{2 \pi} \frac{2i D^{N}}{\omega + i 0^{+}} e^{-i \omega t} \,d\omega
%  = 2 \pi \lim_{\omega \to i 0^{-}} -\frac{\Delta \phi}{2 \pi} D e^{-i \omega t}
\overset{Res}{=} -  2 D^{N} \Delta \phi.
\end{align}
We have calculated the ratio of the peak to long time currents also for finite $\Delta \phi$  and estimate the ratio of the Drude weight intervening in the quench to the sum-rule expectation value:
\begin{align}
\frac{j^{N}(t\to\infty)}{j^{N}(t\to 0^{+})} = \frac{2 D^{N}_{\text{quench}}}{\tau^{N}}. 
\label{eq:quench-ratio}
\end{align}
We stress that the actual stiffness is defined within LR by Eq.(\ref{dtrue}). The results for  $D^{N}_{\text{driving}}$ and 
$D^{N}_{\text{quench}}$ are expected to merge with $D^{N}$
when LR is applicable respectively to a system driven by a nonzero field and a system that relaxed after a nonzero quench of the magnetic flux.    
All these estimates of the stiffness are compared in Fig. \ref{fig7}(b).
For vanishing electric field $D^{N}_{\text{driving}}$ approaches $D^{N}_{\text{Mazur}} \ll D^{N}$, whereas $D^{N}_{\text{quench}}$ nicely reproduces 
the LR result $D^{N}$. The latter agreements holds also for strong quenches $\Delta \phi$, i.e. for relaxation from 
far--from--equilibrium states.
The deeper understanding of the contrasting result for driving and relaxation remains an open problem and requires further studies.
In particular, it remains to be checked whether $j^E(t)/j^N(t)$ approaches $R_{\text{Mazur}}$ also for other driven integrable systems.  

\section{Summary}

We have studied an integrable one--dimensional system of interacting spinless fermions and established the long--time ratio of the energy current
($j^E$) and the particle current ($j^N$)  under dc driving by nonzero electric field $F$.  
The equilibrium LR theory predicts singular (ballistic) responses of both currents, as quantified by the stiffnesses $D^E$ and $D^N$, respectively. Since $j^E$ is a conserved quantity (at $F=0$), $D^E$ represents simply the stress coefficient. However, $j^N$ is not conserved and the physical origin of a finite $D^N$ is more complex. We have first considered a system (doped insulator)  where the local conserved quantities saturate the Mazur bound on $D^N$. In this case the long--time results for $j^E(t)/j^N(t)$ agree with the LR ratio $D^E/D^N$, despite the currents themselves 
are steady or oscillating in contrast to the LR prediction $j^{N(E)} \propto t$.  We have then studied a system 
(doped metal close to half--filling) where large $D^N$ cannot be explained by the Mazur bound formulated in terms of local conserved quantities.
On the one hand, the ratio $j^E(t)/j^N(t)$ obtained for a system which relaxes after a flux--quench ($\delta$--like pulse of electric field) nicely agrees with the LR theory. On the other hand,  $j^E(t)/j^N(t)$ obtained for a steady driving becomes much larger than the LR value $D^E/D^N$.  While the deviation from the LR theory in the latter case is evident, we are not aware of any qualitative explanation for this discrepancy.  
%A conjecture has been formulated in the preceding section but requires further studies.             

\textit{Acknowledgments.} This work has been carried out within 
the project  DEC-2013/09/B/ST3/01659 financed by the Polish National Science Center (NCN). P.P. acknowledges the support by the Program P1-0044 and project J1-4244 of the Slovenian Research Agency.

\bibliography{bibliography.bib}

\end{document}